\documentclass[manuscript]{emulateapj}
\usepackage{times}

\slugcomment{}

\shorttitle{Double red clump kinematics and abundances}
\shortauthors{De Propris et al. }

\begin{document}


\title{Separating the conjoined red clump in the Galactic Bulge: Kinematics
and Abundances.}


\author{
Roberto De Propris\altaffilmark{1}, R. Michael Rich\altaffilmark{2}, 
Andrea Kunder\altaffilmark{1}, \\
Christian I. Johnson\altaffilmark{2}, Andreas Koch\altaffilmark{3}\\
\sc and \\
Sarah Brough\altaffilmark{4}, Christopher J. Conselice\altaffilmark{5},
Madusha Gunawardhana\altaffilmark{6}, David Palamara\altaffilmark{7},\\ 
Kevin Pimbblet\altaffilmark{7}, Dinuka Wijesinghe\altaffilmark{6}
}

\altaffiltext{1}{Cerro Tololo Inter-American Observatory, La Serena, Chile}
\altaffiltext{2}{Department of Physics and Astronomy, University of California, Los Angeles, CA, USA}
\altaffiltext{3}{Department of Physics and Astronomy, University of Leicester, University Road, \\ Leicester LE1 7RH, UK}
\altaffiltext{4}{Australian Astronomical Observatory, Epping, NSW, Australia}
\altaffiltext{5}{Department of Physics and Astronomy, University of Nottingham, UK}
\altaffiltext{6}{Sydney Institute for Astronomy, School of Physics A29, The University of Sydney, NSW, Australia}
\altaffiltext{7}{School of Physics, Monash University, Clayton, VIC, Australia}



\begin{abstract}
We have used the AAOMEGA spectrograph to obtain R $\sim 1500$ spectra of 714 stars that are
members of two red clumps in the Plaut Window Galactic bulge field $(l,b)=0^{\circ},-8^{\circ}$.   
We discern no difference between the clump populations based on radial velocities or 
abundances measured from the Mg$b$ index.  The velocity dispersion has a strong trend with 
Mg$b$-index metallicity, in the sense of a declining velocity dispersion at higher metallicity.  
We also find a strong trend in mean radial velocity with abundance.  Our red clump sample 
shows distinctly different kinematics for stars with [Fe/H] $<-1$, which may plausibly be 
attributable to a minority classical bulge or inner halo population. The transition between the
two groups is smooth. The chemo-dynamical properties of our sample are reminiscent of those 
of the Milky Way globular cluster system. If correct, this argues for no bulge/halo dichotomy 
and a relatively rapid star formation history. Large surveys of the composition and kinematics 
of the bulge clump and red giant branch are needed to define further these trends.

\end{abstract}





\section{Introduction}

The existence of a prominent bar in the Galactic bulge is now well established from
multiple lines of evidence (e.g., \citealt{liszt80,blitz91,stanek94,babusiaux05}). The 
large-scale kinematics of the bulge sampled by the BRAVA (Bulge Radial Velocity
Assay) survey \citep{rich07,howard08,howard09} can be fitted with simple cylindrical 
rotation and little or no classical spheroidal component \citep{shen10}, a conclusion also
reached (albeit at lower confidence) from proper motion surveys \citep{rattenbury07}.
Our `bulge' appears to consist of a peanut-shaped bar, reminiscent of the `pseudobulges' 
defined by \cite{kormendy04} and encountered elsewhere \citep{kormendy10}.

At the same time, this result is puzzling. The BRAVA data imply that our Galaxy has not 
undergone any significant merger since the epoch at which the disk formed, in contrast 
with expectations from simulations within $\Lambda$ CDM cosmological models (e.g., 
\citealt{bullock05,cooper10}). However, it is well known that stars in the bulge are 
metal-rich and $\alpha$-enhanced, indicating a rapid star formation history typical of 
classical bulges \citep{mcwilliam94,ballero07}. The observation of an abundance gradient 
for bulge stars \citep{zoccali08} and a correlation between abundance and kinematics 
\citep{babusiaux10} may also indicate the presence of a classical spheroidal component, 
although dynamical data \citep{shen10} seem to be inconsistent with this . On the other
hand, detailed abundances for subgiants and lensed dwarfs, are more similar to those of 
the thick disk, in accordance with a pseudobulge formation scenario \citep{melendez08,
alves10,bensby10,ryde10}. 

Adding to the complexity, the red clump in the bulge color-magnitude diagram appears doubled 
at $|l| > 5^{\circ}$ \citep{mcwilliam10,nataf10}. \cite{mcwilliam10} carry out a very careful analysis 
of the photometric properties of the double red clump and conclude that the splitting of the red 
clump is due to a distance effect and that the bulge may contain an X-shaped structure, extending 
from the ends of the bar. So far, the only spectroscopic study of this population has been carried 
out by \cite{rangwala09a} using Fabry-Perot imaging spectroscopy in Baade's Window and some 
adjacent fields, and their results suggest that there is a metallicity gradient with galactic 
latitude \citep{rangwala09b}, as well as dynamical differences, which would be more 
consistent with the presence of a stream, although it is clear that more accurate kinematics 
and metallicities are needed to truly understand the nature of this feature.

Here we present the first low resolution survey of a field at $l=0^{\circ}$ $b=-8^{\circ}$
(Plaut's Window) specifically targeting the red clump stars. We analyze the kinematics and 
metallicity of stars belonging to each peak and derive the metallicity-kinematics trends. 
Although this is clearly a `first look' exploratory analysis, our data imply that the two populations 
are dynamically and chemically similar and favor the X-shaped bulge hypothesis 
proposed by \cite{mcwilliam10}.

\section{Data Reduction and Analysis}

The data for this project were kindly provided by the Galaxy and Mass Assembly
Survey (GAMA -- \citealt{driver10}). This survey is observing three 12 deg$^2$
fields at 9h, 12h and 15h RA using the AA$\Omega$ multi-fiber spectrograph on
the Anglo-Australian Telescope (AAT). At the end of May, 2010, it was
found to be impossible to reach either GAMA field during the last two hours of
the night, at sufficiently low airmass. The survey kindly offered to observe one
of the fields containing a double red clump for us. 

Two 400-fiber configurations were observed for Plaut's Window (where the two red clumps
are relatively well separated and the extinction is lower). The target stars were selected from 
the 2MASS survey \citep{skrutskie06}, in a 2 degree (diameter) field.  Bulge stars were required
to have $K_0=7.5\,(J-K)_0+9$ to exclude disk contamination and to lie between
$12.5 < K_0 < 13.5$ and $0.58 < (J-K)_0 <  1.00$ to probe the double red clump.
Magnitudes were dereddened from the \cite{schlegel98} maps. Figure 1 shows 
the color-magnitude diagram of the target stars. The two red clumps are identified
by applying a Gaussian Mixture Modelling algorithm to the data \citep{muratov10}, 
which also assigns to each star a probability of belonging to either peak. The magnitudes 
of the two red clumps peaks are in good agreement with those reported by \cite{mcwilliam10} 
for this region. By necessity, we used the same spectroscopic setup as employed for the 
GAMA survey; this covers the entire optical window, from about 3700 to 9000 \AA\ at a 
resolution of about 1500. Although this is somewhat less than optimal for determining stellar
radial velocities and abundances, it suffices for our initial analysis of this field.

\begin{figure}
\epsscale{0.9}
\plotone{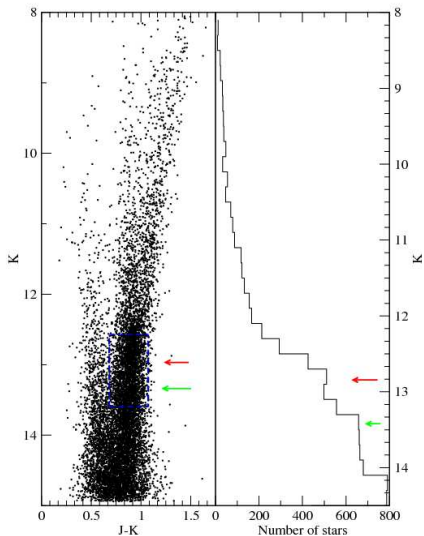}
\caption{The color-magnitude diagram (from 2MASS data) for our bulge
field is shown in the left-hand panel of this figure. The blue box shows
the (non-dereddened) selection limits used. The right-hand panel
shows the histogram of the distribution of stars in $K_0$ and the two
peaks identified by the GMM algorithm, marked by arrows (red for the
brighter peak and green for the fainter peak).}
\end{figure}

The data were reduced using the automated pipeline supplied by AA$\Omega$. A total of 714 
stars were eventually observed. The typical signal to noise ratios of these spectra are about 10 
at Ca H \& K, 20 in the Mg$b$ region and 50 for the Calcium Triplet wavelength range. Radial 
velocities were measured by cross-correlation against stellar templates using the {\tt runz} 
program \citep{saunders04} which is specifically written for analysis of AA$\Omega$ data. Among
the several templates available in {\tt runz} we imposed the choice of a K-giant template, if this was 
not automatically selected by the program. Of the 714 targets, 631 returned a valid radial velocity
($> 95\% $ probability that the measured velocity is correct from the height of the correlation peak).
Typical velocity errors are 1/10 of a resolution element, or 25 km s$^{-1}$ for the $R \sim 1500$ 
of GAMA data, but no stars with a velocity error above 2$\sigma$ (50 km s$^{-1}$) were used 
for our analysis.

We experimented with a number of techniques to measure metal abundances, including the 
non-SEGUE  stellar parameters pipeline (e.g., \citealt{li10} and references therein) and the
calcium triplet (although red clump giants were outside the luminosity range of the calibrations). 
We eventually found that the most reliable measurements (when compared with the high 
resolution abundances measured for giants in Plaut's window by \citealt{johnson10}) were given 
by using the Mg$b$ Lick index \citep{worthey94a, ibata95}. We measured this index and its
errors  (based on the CCD noise parameters and the radial velocity errors) using the 
{\tt LECTOR} software\footnote{http://www.iac.es/galeria/vazdekis/vazdekis\_software.html}. 
The  typical error in measuring the Mg$b$ index is about 5\%.

In order to derive metal abundances from the Mg$b$ index, we used fitting functions for
cool stars by \cite{worthey94b}. We assumed $\log g = 2.25$ which is appropriate for
red clump giants based on the Padova isochrones \citep{marigo08}. Temperatures were
estimated from the dereddened $J-K$ color, using the $T_{eff}$ -- color relations 
by \cite{houdashelt00}. This approach was used by \cite{cote97}, among others, to
measure the abundance of red giants in M31, although we found that only Mg$b$ can
be reliably measured in our data. We derived metallicities for 545 stars in our data. 

The typical random error in metal abundance is $\pm 0.1$ dex, based on the error in
the index measurement, but of course there are systematic errors depending on the 
assumed values of $\log g$ and $T_{eff}$. Altering $\log g$ by 0.1 at the same temperature 
yields a change in [Fe/H] of around 0.15 dex, while altering T$_{eff}$ by 100K yields abundance 
changes of 0.3 dex. In addition, our data are not on the Lick system as no standards were 
observed, and this may introduce a systematic effect of the order of 0.2 dex in the measurement 
of metal abundances. We used some Lick standards observed by ELODIE \cite{moultaka04}
to measure Mg$b$ and find that our abundances tend to be $\sim 0.2$ dex too low, which
would bring our data in better agreement with high resolution measurements by \cite{johnson10}.
In addition, the known $\alpha$ enhancement of the bulge (e.g., \citealt{mcwilliam94})
means that the Mg$b$ abundance may overestimate the actual [Fe/H] of the stars, although
it is probably a better proxy of the total metal abundance. 

\section{The nature of the red clump}

Figure 2 shows the radial velocity distribution for red clump stars derived from our data.
The mean heliocentric radial velocity for the entire sample is $-15 \pm 5$ km s$^{-1}$
(rms) , which is equivalent to a velocity of $-4$ km s$^{-1}$ (Galactocentric standard of 
rest), consistent with data from the BRAVA survey. The radial velocity dispersion is $110
\pm 5$ km s$^{-1}$, which is somewhat larger than what is measured for this galactic latitude
by BRAVA, although we are looking at a fainter sample. The intrinsic dispersion and its error
were computed following \cite{spaenhauer92}. The higher velocity dispersion seems to be 
due to a low metallicity population that may not be present in the BRAVA data.

\begin{figure}
\plotone{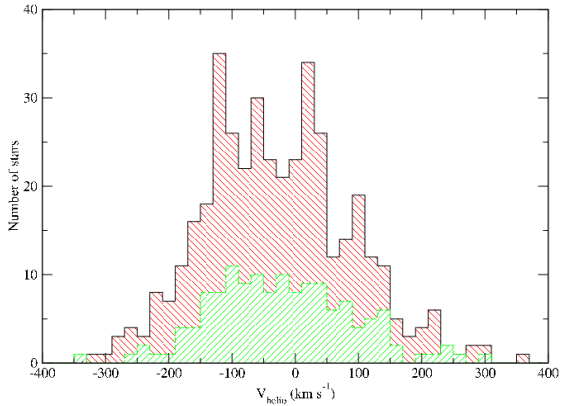}
\caption{Radial Velocity distribution of stars belonging to each red clump peak (red 
for the brighter peak and green for the fainter peak as in Figure 1).}
\end{figure}

Stars in the first (brighter) peak have a mean heliocentric velocity of $-18 \pm 6$ km s$^{-1}$ 
and velocity dispersion of $109 \pm 5$ km s$^{-1}$, while stars belonging to the second (fainter) 
red clump have mean velocity of $-6 \pm 8$ km s$^{-1}$ and velocity dispersion of $113  
\pm 9$ km s$^{-1}$.  Within the errors, stars in both red clumps have the same mean velocity 
and velocity  dispersion and therefore appear to have the same kinematics. This would appear 
to be at odds with claims for a difference in the kinematics of the two red clumps 
\citep{rangwala09a}, although more sightlines are needed. Application of the Gaussian 
Mixture Modelling algorithm shows that there is $<50\%$ chance that the velocity distribution  
is bimodal. A Kolmogorov-Smirnov two-sample test yields an 88\% chance that the two 
distributions come from the same parent. The t-test and F-test show that the two distributions
do not have significantly different means or variances.

Figure 3 shows the distribution in metal abundance for stars in both clumps. The distribution
for all stars has a mean [Fe/H] of $-0.61 \pm 0.03$. The distribution is consistent with that observed
for red giants in Baade's Window by \cite{mcwilliam94}, the high resolution metallicities
in Plaut's window (overplotted in the figure) measured by \cite{johnson10} and the expectations
from the metallicity gradient measured by \cite{zoccali08}, but with a systematic offset 
($\sim 0.2$ dex)  probably due to the lack of an absolute calibration based on Lick standards.

\begin{figure}
\plotone{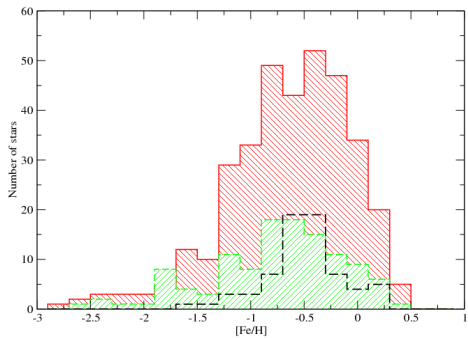}
\caption{[Fe/H] distribution of stars belonging to each red clump peak (red for the
brighter peak and green for the fainter peak as in Figure 2). The black dashed 
histogram shows the distribution of metal abundances for red giants studied by
Johnson et al. 2011.}
\end{figure}

Stars in the first peak have mean [Fe/H] of $-0.55 \pm 0.03$ with a dispersion of $0.58$, 
while stars in the second peak have mean [Fe/H] of $-0.67 \pm 0.03$ with a dispersion of 
$0.62$. Again, the Gaussian mixture modeling algorithm returns a distribution consistent 
with a unimodal distribution. Within the errors, stars belonging to each red clump have 
the same [Fe/H] abundances. A K-S test gives a 73\% probability for the two clumps to be 
drawn from the  same distribution. Similarly, the t-test and F-test show that two samples 
do not have significantly different means or variances.

One aim in this work is to compare the kinematics and metallicities of stars belonging
to the two red clumps identified in the galactic bulge by \cite{mcwilliam10,nataf10}. 
Our data show that the two populations have the same kinematics and metal abundance, 
which suggests that the difference in luminosity between the two red clumps is a distance 
effect.  In other words, the two red clumps are observed at the two ends of the $\sim 2$ kpc 
bar that constitutes the bulge of the Milky Way. These data are therefore consistent with the 
analysis by \cite{mcwilliam10}, attributing the double red clump to the existence of X-shaped 
protrusions at the end of the Milky Way bar (i.e., an X-shaped bulge).  If these stars lie at the 
end of a bar, then the metal abundance gradient across the structure is expected to be quite 
small.

An alternate possibility is that the two red clumps are different in helium content.
Helium abundance variations are believed to exist (from indirect evidence) in massive
globular clusters in our Galaxy (e.g., \citealt{carretta09}). Helium enhancement would 
produce differences in the luminosity of the red clump stars \citep{dantona10}, but any
process capable of producing the necessary helium enhancement would also overproduce
metals. We would presume that populations with different chemical abundances might also
have different kinematics. 

\section{Abundance trends with kinematics}

Classical bulges present a number of abundance trends with kinematics, some of
which are also observed in the Galactic bulge. \cite{zoccali08} have measured a 
metallicity gradient with radius. \cite{babusiaux10} find a trend between metal abundance 
and velocity dispersion in Baade's Window and two lower ($b=-6^{\circ}$ and $-8^{\circ}$)
galactic latitude fields, arguing for a two component bulge, with a metal-rich system
comprising the bar and a metal-poor spheroid or thick disk, the relative contribution from 
each of these varying with galactic latitude.

We plot the mean velocity (lower panel) and velocity dispersion (upper panel) as a function 
of metallicity in Figure 4. We plot all stars (black points), stars in the first (brighter)
peak (red points) and stars in the second (fainter) peak (green points) separately.
Stars in each group are binned in bins containing the same number of stars after
sorting by metal abundance. For each of the groups we consider (all, stars in the
brighter and fainter peaks) the bins span a non-overlapping range in metal abundance. 
Stars in both red clumps appear to obey the same relations between metallicity and 
kinematics. This confirms that the two populations do not differ in kinematic properties 
or abundances, which is more consistent with a projection effect and an X-shaped bulge 
structure.

\begin{figure}
\plotone{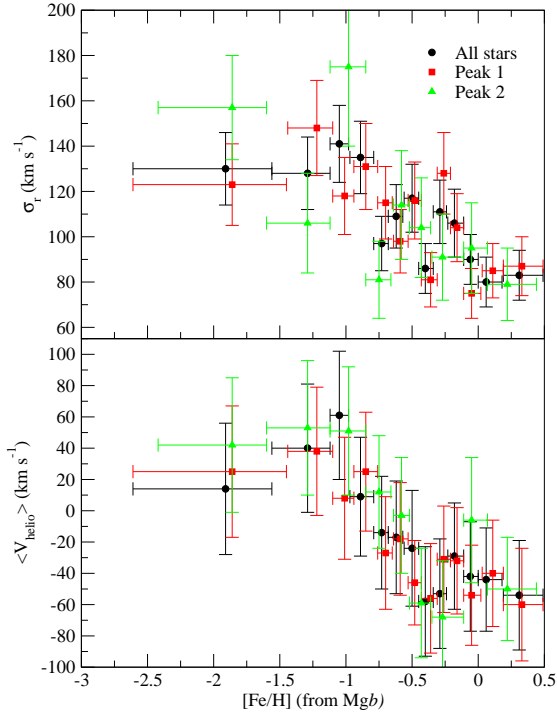}
\caption{Dependence of heliocentric mean velocity on metallicity (lower panel) and
radial velocity dispersion (upper panel) for stars in our field. See legend in figure to
identify the samples. The bins are chosen to contain the same number of stars, sorted 
by metal abundance; the horizontal error bars represent the range of metallicities in each 
bin. The vertical error bars are the errors on mean radial velocity and velocity dispersion, 
as appropriate.}
\end{figure}

The data shown in Figure 4 show a clear trend for increasing velocity dispersion with
decreasing metal abundance. This is similar to what is found by \cite{johnson10} in their
higher resolution data.  \cite{babusiaux10} uses two fields at $b=-6^{\circ}$ and $b=
-12^{\circ}$ and although there are few stars at [Fe/H] $< -1$ in their data, there is a
hint of an increase in the velocity dispersion at lower metal abundances, in agreement
with our observations. \cite{vieira07} find a flat distribution of metallicity with velocity
dispersion (from proper motion data) for stars with [Fe/H] $> -1$, which is not in
disagreement with our observations, where most of the increase in velocity dispersion
takes place for lower metallicity stars.

For stars with [Fe/H] $ > -1$ the velocity dispersion is in good agreement with that
measured for BRAVA M giants and does not depend strongly on metal abundance,
which is broad agreement with what measured by \cite{vieira07} in Plaut's Window
and the two lower galactic latitude fields in \cite{babusiaux10}. However, at [Fe/H] $<-1$
there appears to be a (at face value) smooth transition to a dynamically  hot component. 
Similarly, in the upper panel of Figure 4 we see that the metal-rich component appears to 
have significant mean heliocentric velocity, while at [Fe/H] $< -1$ one observes a smooth 
trend towards a relatively static velocity component. One caveat in this is that the metallicity
errors are large, and we cannot rule our a bimodal distribution with the `wings' of the errors 
simulating a smoother transition between the two behaviors, although this would require 
a correlation between metallicity errors, measured radial velocity and velocity dispersion.

The metal rich stars are best interpreted as part of the bar/bulge structure. Their
kinematics show evidence of rotational support and bulk rotation and are 
consistent with data from the BRAVA survey in this region. The behavior of the
more metal poor component, showing high velocity dispersion, and low to
zero velocity relative to the Sun may be explained by a classical bulge or by
inner halo stars. With a mean metal abundance of [Fe/H] $\sim -1.5$ these
stars appear to be best interpreted (at least provisionally) as an inner halo population,
although bulges can of course be metal poor as well. This is consistent with the
observations by \cite{zoccali08} and \cite{babusiaux10}, albeit for a single
sightline. However, the BRAVA data show no classical bulge component fitting
their dynamical model \citep{shen10}. One possibility is that by selecting M
giants and using the Calcium Triplet as their main radial velocity indicator,
BRAVA may be biased against lower metallicity stars and therefore preferentially
miss the high $\sigma$ component.

The properties of galactic globular clusters present an interesting analogy with what is
observed here: metal rich clusters, with mean [Fe/H] of $\sim -0.7$ are believed to be
associated with the bulge and are supported at least in part by rotation, whereas inner 
halo clusters have mean [Fe/H] of $\sim -1.6$ their kinematics are dominated by random
motions and at most very slow rotation. It is tempting to speculate that the two components 
we see in our data are analogous to the metal-poor and metal-rich globular clusters, whose
properties they appear to share to some extent (cf., \citealt{babusiaux10} for a similar 
two-component model for the bulge). \cite{ortolani95,zoccali03} and \cite{clarkson08} have 
argued, on the basis of isochrone fits to bulge globular clusters and field stars, that the
bulge formed nearly coevally with the halo. Most globular clusters in the inner halo 
formed within $\pm 1$ Gyr of each other \citep{marin09}. If this applies to bulge stars
as well, it would imply a rapid star formation process, at least for the inner regions 
($ < 20$ kpc) of the Milky Way

If this is the case, the smooth transition between the metal-rich and metal-poor subsystems, 
with a `turn-over' point at [Fe/H] $\sim -1$ may imply that the bulge and halo components 
are continuous and that there is no clear dichotomy between the two (modulo the large 
errors in metal abundance). This would be consistent with the BRAVA result that the bulge 
was formed (in a dynamical sense) from secular evolution at high redshift. As long as the 
stars also formed rapidly, the observed $\alpha$-element enhancements are not in
disagreement with this hypothesis.

\acknowledgments

This publication makes use of data products from the Two Micron All Sky Survey, which is a joint
project of the University of Massachusetts and the Infrared Processing and Analysis Center/
California Institute of Technology, funded by the National Aeronautics and Space Administration
and the National Science Foundation. R. Michael Rich acknowledges support from grant AST 
0709479 from the National Science Foundation.  We would like to thank the GAMA survey for
allowing us to use two hours of AAT time to carry out this program. We would also like to
thank Scott Croom, David Wilkerson ,Rob Sharp and Will Sutherland.



{\it Facilities:} \facility{AAT (2dF)}.

\end{document}